# Plasma propagation dynamics in a patterned dielectric barrier discharge at different length scales

Zaka-ul-islam Mujahid[1] and Julian Schulze[1,2]

1. Department of Electrical Engineering and Information Science, Ruhr-University Bochum, D-44780, Bochum, Germany
2. Key Laboratory of Materials Modification by Laser, Ion, and Electron Beams (Ministry of Education), School of Physics, Dalian University of Technology, Dalian 116024, People's Republic of China

**Abstract**

Packed bed plasma reactors (PBPRs) inherently have complex geometries where the volume between the electrodes is filled with dielectric/catalytic pellets to form an array of voids and cavities. This design generates multiple length scales i.e. the dimension of the plasma region (of the order of centimeters), of the dielectric/catalytic pellets (of the order of millimeters) and of the void between the dielectrics (or cavities) and/or surface pores that can reach micrometer dimensions. The plasma is formed as an array of connected or segmented microdischarges. The understanding of plasma propagation on these diverse length scales is important in optimizing and controlling this process. In this work, we investigate the discharge formation and propagation at different length scales as a function of applied voltage in a simplified design of PBPRs, a patterned dielectric barrier discharge (p-DBD), operated in helium at atmospheric pressure, using phase and space resolved optical emission spectroscopy (PROES). The p-DBD is based on an array of plasma facing glass hemispheres implemented into one of two electrodes and regularly arranged in a hexagonal pattern. In this way the stability of and the diagnostic access to the plasma are improved. The plasma is initiated as filamentary microdischarges at the positions of minimum gap. This phase is followed by radial acceleration of electrons over the dielectric surface as surface ionization waves to finally form surface microdischarges at the voids between the pellets, i.e. at the contact points. With the increase of the voltage amplitude, subsequent pulses of surface microdischarges are generated at the contact points. The plasma emission of each of the surface micro-discharges (S-MD) is found to consist of microscopic structures whose emission intensity increases as a function of the driving voltage amplitude. Adjacent microdischarges interact to generate a wave-like emission intensity propagation from the center of the array to the edges.

1. **Introduction**

Non-thermal plasmas such as dielectric barrier discharges (DBDs) are widely investigated for multiple environmental and biomedical applications [1; 2]. Packed bed plasma reactors (PBPR) in particular are an attractive option for gas reprocessing applications such as gaseous pollutant removal [3] and greenhouse gas valorization [4]. A PBPR is a dielectric barrier discharge (DBD) in which the volume between the electrodes is filled with catalyst (dielectric) pellets. The combination of plasma and catalyst has been reported to work in a synergetic manner [5]. However, this results in a complex layout of the plasma source with a multitude of design as well as operational conditions and catalytic parameters which are difficult to understand, predict and control [6]. The situation is more complex than the discharge formation in a single filament [7] or over the surface of a single pellet [8]. From the design aspect, packing the volume DBD with dielectric pellets (often porous) generates an array of connected cavities with multiple length scales where plasma is generated. More precisely, the length of the plasma reactor is typically of the order of centimeters, which corresponds to the macroscale, the size of the pellets is typically of the order of mm and corresponds to the meso scale and the size of the cavities between the dielectric pellets and of the surface pores can reach µm (microscale). Whitehead [5] has recently reviewed the current progress in understanding both plasma and catalyst aspects at multiple length and time scales and suggested further improvement in performance requires understanding the process across these scales.

Previous results of time resolved quasi-3D experimental imaging and 2D models of such plasma sources driven by sinusoidal voltage waveforms with amplitudes of several kV showed an excellent qualitative agreement [9]. It was found that plasma is generated due to three distinct mechanisms which generate distinct current pulses. In a PBPR the discharge is initiated as a cathode directed positive streamer which is converted into a filamentary microdischarge (FMD) during the first discharge current pulse. This is followed by the radial acceleration of electrons over the surface of the dielectric as surface ionization waves (SIWs) (during the second current pulse) which are consequently transformed into a surface microdischarge (SMD) after reaching the region between the dielectrics, i.e. the contact points. Depending on the voltage amplitude, several additional discharge current pulses are generated which produce SMD at the contact points. A similar combination of filamentary microdischarges, surface ionization waves and surface microdischarges has been observed in a different geometry [10].

Experimental results showed that the discharge does not penetrate into nanopores (pore size<0.8 µm) but penetrated into mesopores (pore size≥15 µm) [11]. Zhang et al. [12; 13] modelled the plasma generation in a pore on a dielectric surface. They observed that ionization, electric fields and electron densities can be stronger inside the pores compared to the bulk, and the applied voltages, pore dimensions and shape are critical parameters. The plasma can only be produced inside the pores if the pore diameter is larger than the Debye length. They showed that at higher voltages ionization can take place mainly inside the pores.

The plasma appears to be formed as segmented into multiple microdischarges [14] or connected [15] or covering the surface [16] depending on the conditions. Fast time resolved diagnostics have shown that the time resolved plasma propagation is quite complex and could not be observed from still photography [9]. In terms of modelling, a single length aspect such as pores or a segment of the reactor have been investigated by 2D models [17]. The more accurate 3D investigation of multiple length scales in a single design could not be performed due to computational limitations. However, the investigation of these multiple length aspects simultaneously is required to weigh their relative role and importance.

Arrays of microdischarges have been observed to show interactions between adjacent cavities which result in wave-like emission intensity propagation in adjacent cavities [18-20]. Numerical investigations showed that the initial plasma (produced inside a given cavity) generated photoemission of electrons in adjacent cavities. Consequently, this results in a drift of electrons followed by a drift of ions between the cavities, which generated a wave-like propagation of emission [20]. A similar behavior could be present in PBPRs. Detailed insights into this phenomenon in PBPRs would result in improved understanding, prediction, optimization and control of the plasma and its applications. Therefore, such interactions between the adjacent microdischarges should be investigated.

Conventional PBPRs have a complex design with limited optical access where the dielectric pellets are typically packed in a random arrangement which makes it difficult to probe the discharge for fundamental understanding using optical emission spectroscopy [16]. An alternative to the conventional PBPR is a patterned dielectric barrier discharge (pDBD) design [9;

21], where instead of irregular filling of the gap with dielectric pellets, one of the dielectrics, which cover the electrodes, is patterned or filled with hemispherical or semispherical pellets. Such a design allows to perform investigations of the space and time resolved plasma dynamics in the complex void volume between the pellets using a combination of quasi 3D diagnostics. It also results in improved plasma stability, which is the basis for applying diagnostics that average data over multiple plasma pulses such as Phase Resolved Optical Emission Spectroscopy (PROES).

In this work, we use small and realistic dimensions in a helium pDBD at atmospheric pressure and driven by sinusoidal voltage waveforms to study the plasma propagation at different length scales by imaging the complete reactor as well as a zoomed region. The results suggest that the plasma is generated as filamentary microdischarges in the volume, surface ionization waves over the surface and surface microdischarges at the contact points of the dielectric structures. Compared to previous work [9] the reduction in dimensions results in surface ionization waves right after the FMD and in the same discharge current pulse. In the complete reactor, there is strong interaction between discharges at adjacent cavities which generates a wave-like plasma propagation from the center to the edges. Within a single void, the surface microdischarge shows multiple sharp emission structures which get more pronounced as a function of the applied voltage.

2. **Experimental setup**

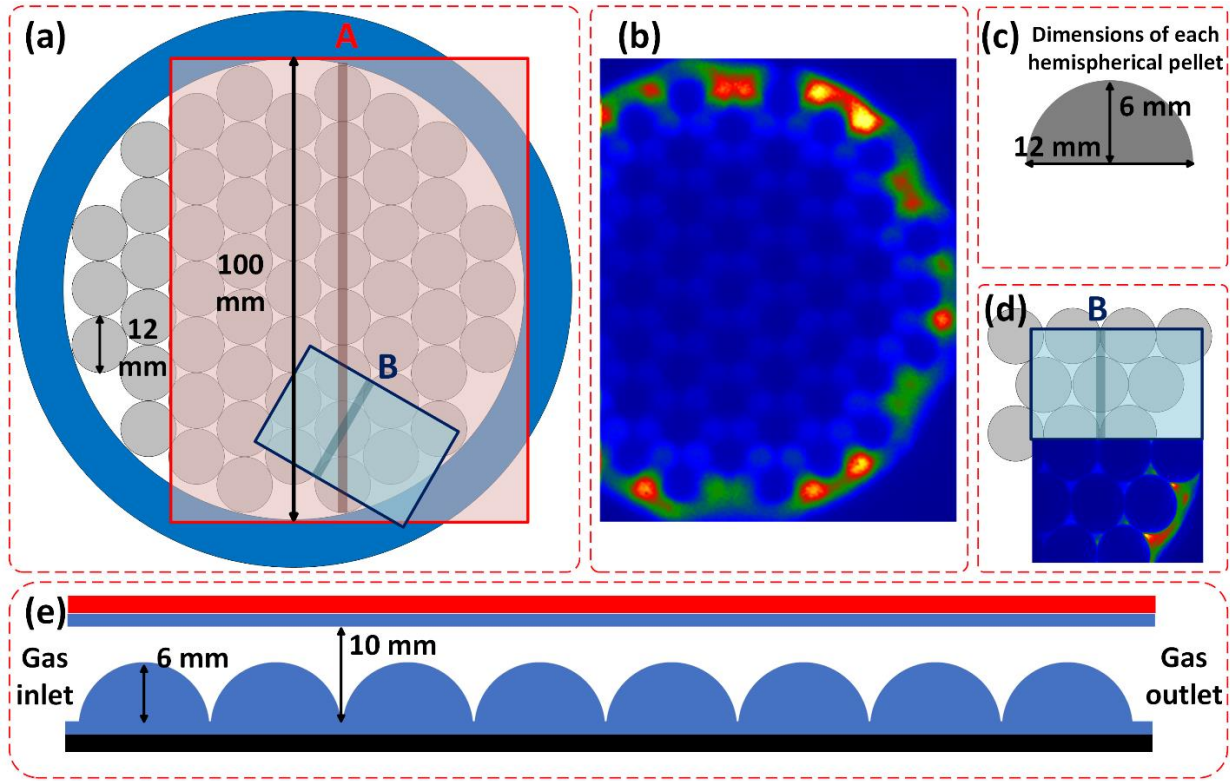

*Figure 1 (a) Schematic of the top view of the reactor with regularly arranged fifty-five pellets (12 mm diameter and 6 mm height) inside a 100 mm diameter cylindrical cell. Two different areas are indicated in red (full view) and blue (zoomed). The vertical line, indicated with A, represents the position where the time resolved emission is analyzed for the full view. The blue bar marked as B indicates this position inside the zoomed region of interest. (b) Typical full view image of the region shown in (a). (c) Dimension of a single hemispherical pellet used in this work. (d) The imaged region in the zoomed view and a representative image are shown. The line B shows the position where time resolved emission is analyzed for the zoomed view. Both representative images for the full view (b) and for the zoomed view (d) were taken at a voltage amplitude of 2.05 kV at 2 µs. (e) Schematic of the sideview of the reactor (not to scale) at the position A with eight pellets linearly arranged.*

The conventional PBR geometry (where the space between two coaxial cylinders is filled with spherical dielectric beads) does not allow performing meaningful optical diagnostics due to limitations in optical access and random generation of microdischarges in a randomly filled space. Therefore, a simplified geometry, the so-called patterned DBD (p-DBD), with regularly packed dielectric pellets is used. A schematic top and side view of the reactor are shown in Figure 1 (a) and (e), respectively. The reactor is a parallel-plate-like DBD where one surface consists of a layer of dielectric hemispheres to incorporate spherical boundaries between pellets and contact points where the pellets meet. A p-DBD allows easier optical access through a transparent electrode at the top. The p-DBD design enables stable and reproducible discharge generation, where space and time resolved diagnostics can be performed. Further details on the p-DBD system are discussed elsewhere [9; 21]. The aim of this work is to understand the plasma propagation in PBPR like systems at different length scales. For this reason, a larger number of smaller dielectric pellets has been used.

The current DBD system (schematic shown in Figure 1 (a)) is generally similar to the one used in previous work [9; 21]. The reactor wall corresponds to a quartz ring with 100 mm diameter and 10 mm height. At the top and bottom of this ring an electrode is placed, respectively. The top electrode corresponds to a glass plate that is coated with transparent and conducting ITO from the outside and serves as powered electrode. Semi-spherical glass pellets, which have a diameter of 12 mm and are 6 mm thick, are embedded into the plasma facing side of the grounded electrode and are arranged in a hexagonal arrangement (see Figure 1 (c)). Compared to previous work, where larger structures were used [9; 21], the smaller size allows to realize a symmetrical arrangement of fifty-five hemispherical pellets in a hexagonal pattern as shown in Figure 1 (a). The maximum gap between the planar dielectric covering the powered electrode and the patterned embedded dielectric structures at the grounded electrode is 10 mm. In order to observe the dynamics at different length scales, two different imaging areas for the full view and a zoomed region are used. These regions are indicated by red and blue transparent rectangles, respectively, in Figure 1 (a). Representative images from the two regions are shown in Figure 1 (b) and Figure 1 (d), measured for voltage amplitude of 2.05 kV at 2 µs. Figure 1 (a) indicates the positions where time resolved changes in emission have been determined for the full and the zoomed view indicated as vertical lines A and B, respectively. The lines A and B have a 30º angle between them and represent a distance of 9.6 and 2.08 cm. Plasma is generated using 2 slm (standard liters per minute) flow rate of helium gas (99.999% purity) at atmospheric pressure and a 10 kHz high voltage power supply that generates a sinusoidal waveform, whose amplitude is varied from 1.8 kV to 5.2 kV. The He 706.5 nm emission from selected plasma regions is measured with 1 µs time resolution (except Figure 7 with 0.1 µs) using an Andor ICCD camera. The He 706.5 nm (He $3^3S$ to $3^3P$) line is spectrally discriminated using a 710 nm interference filter with a bandwidth of 10 nm. This line is mainly populated by electron impact excitation from the ground state and other mechanisms are negligible [22]. Therefore, the modulation of the 706.5 nm line emission generally reflects the dynamics of energetic electrons in response to the applied voltage. The ICCD camera is operated in a phase synchronized manner with the applied voltage signal using an external signal generator with an internal delay generator to realize Phase and space resolved optical emission spectroscopy (PROES). Further details on the PROES setup can be found in Refs. [23-25].

## 3. Results

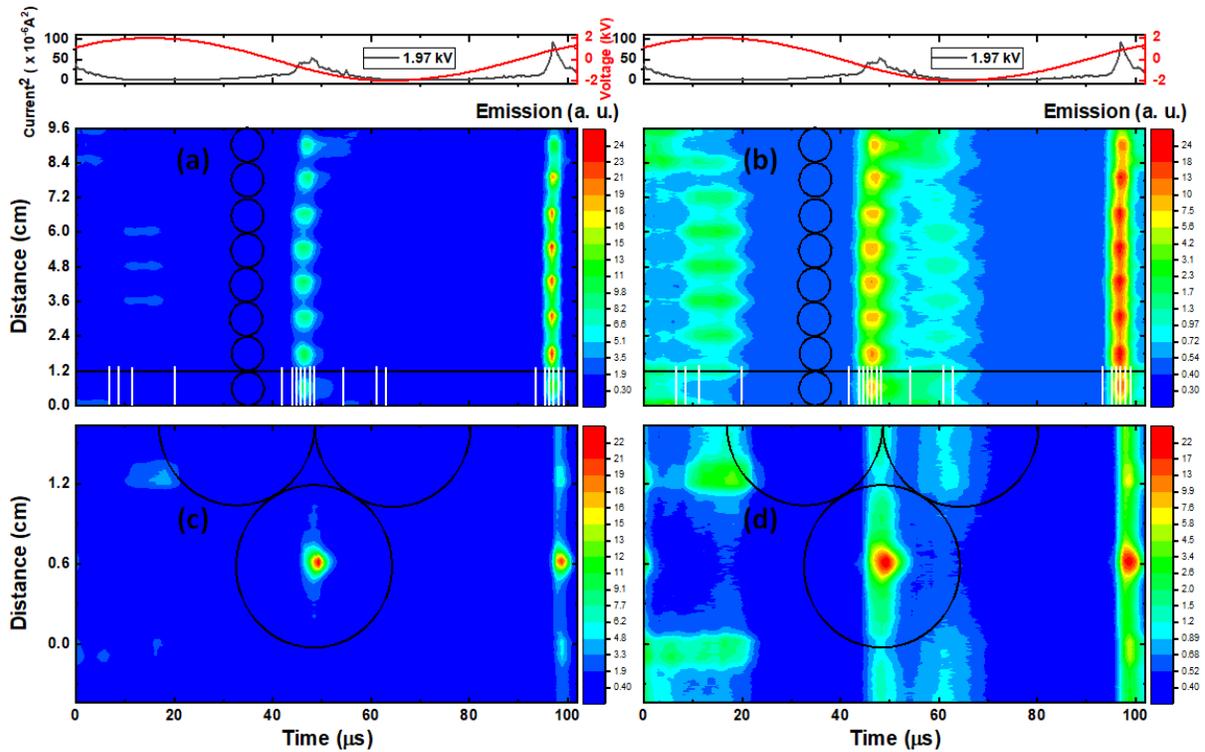

Figure 2: Spatio-temporally resolved optical emission from the p-DBD operated in helium at atmospheric pressure with a sinusoidal voltage waveform (10 kHz, 1.97 kV amplitude). The horizontal axis covers one period of the applied voltage waveform. (a, b) show the space and time resolved emission at the position A in the full view as indicated in Figure 1 (a). The white lines indicate the temporal position of the 2D images shown in Figure 3. Spatial resolution is provided along the bar indicated in Figure 1 (a) inside the red region of interest. The relative position of eight 1.2 cm diameter pellets is shown as circles. (c, d) shows the time resolved emission from the position B in the zoomed view as specified in Figure 1 (d). Spatial resolution is provided along the bar indicated in Figure 1 (a) inside the blue region of interest. The positions of the central and two upper pellets are shown as circles. The typical space-time contours plotted using a linear scale are shown on the left [(a) and (c)]. The right side (b, d) shows the same space time contours plotted on log scales. The emission structures at ≈ 44 µs and ≈ 95 µs correspond to the negative and positive polarity at the powered electrode, respectively. The associated measured time resolved current square and voltage waveforms are shown at the top in both columns.

Figure 2 (a-d) shows the time resolved emission from the discharge at 706.5 nm for an applied voltage amplitude of 1.97 kV. The horizontal axis shows the time duration (102 µs) of one complete period of the driving voltage waveform and the vertical axis provides spatial resolution in the direction parallel to the electrodes at position A (a, b) and B (c, d) according to Figure 1 (a). Measurements were taken from the top through the transparent electrode so that no spatial resolution perpendicular to the electrodes is available. The position A is in the center of eight linearly arranged 1.2 cm diameter hemispherical pellets. The position of these eight pellets is indicated by circles in Figure 2 (a) and (b). Figure 2 (c) and (d) show the time resolved emission from the zoomed region measured at the position B, indicated in Figure 1 (a, d). Due to significantly lower intensity, some emission structures are not clearly visible on the left if the typical linear scale is used. Using the logarithmic scale on the right, all emission structures can be simultaneously observed. For similar reasons, all the following space time contours are plotted on logarithmic scales. The measured time resolved current square and voltage waveforms are compared in both columns.

Figure 3 shows selected 2D space resolved images of the time resolved emission shown in Figure 2 (b). Each image is labelled with the time of acquisition and is normalized to its maximum to clearly illustrate the emission structure at that time. The relative changes of the emission are shown by the time resolved contour plot in Figure 2 (b). The respective times of acquisition within the RF period of the 2D images in Figure 3 are indicated by white lines in Figure 2 (a, b).

Figure 2 (b) shows that the emission is initiated at the time of positive and negative voltage polarities at ≈ 44 µs and ≈ 95 µs. The first emission structure in both half cycles has eight maxima at different axial positions. Maximum emission is generated in the center of each of the eight dielectric structures i.e. at the location of minimum gap between the dielectrics at the powered and grounded electrode. The zoomed view in Figure 2 (d) indicates that the emission is maximum in the center but it spreads up to the end of the dielectric pellet position or contact point in both halves. The spread of the emission maxima to the edge of the pellet indicates radial acceleration of electrons.

The first emission peak is followed by a secondary weaker emission structure with maxima at ≈ 60 µs and ≈ 15 µs, which occurs at different positions. The position of the emission maxima with respect to the position of the glass pellets (indicated by the horizontal line in Figure 2 (b)) shows that the secondary emission maxima are generated at the contact points of the dielectric hemispheres.

Previous comparisons of quasi-3D PROES measurements and 2D model results with larger pellets [9] showed that the discharge is initiated at the position of minimum gap as cathode directed positive streamer in both halves which generates weak emission. During the positive half period the streamer head propagates towards the structured dielectric and during the negative half period it propagates towards the planar dielectric. Once the streamer reaches the opposite side and the electrode gap is conductive above the center of the dielectric hemispheres a filamentary microdischarge (FMD) is generated in the center of the pellet with relatively strong emission. The FMD is eventually extinguished once the dielectric at this position is fully charged. Such FMD discharges are followed by so-called surface ionization waves (SIW), where a second emission structure and discharge current peak are generated due to radial acceleration of electrons over the surface of the dielectric. Depending on the instantaneous polarity of the driving voltage waveform, the positive streamer head either hits the structured or the planar dielectric, which correspond to the instantaneous cathode, where the dielectric charges up positively. Once the positive streamer reaches the dielectric, electrons are accelerated from the sides towards the streamer head. In this way, ionization is caused on the sides of the impingement position and the streamer propagates laterally, i.e. SIWs are formed, which propagate from the center towards the contact points of neighboring dielectric structures. At the opposite electrode (instantaneous anode) electrons hit the dielectric, which charges up negatively. In both halves of the pulse period, however, the spatial evolution of the plasma is determined by the structured dielectric surface at the grounded electrode.

The results obtained here are similar to this previous work except the fact that the initial emission maximum does not extinguish in the center, but spreads to the edge of the dielectric pellet and reaches the contact points. The spread of this emission maximum indicates acceleration of electrons tangent to the surface of the dielectric. Therefore, the FMD and SIW are combined in a single emission structure and discharge current peak. This is because smaller dielectric pellets with higher curvature are used here, which are expected to generate stronger electric fields due to polarization of the dielectric pellets. The emission intensity is maximum in the center of the dielectric structure and almost ten times weaker during the radial acceleration. The secondary emission structures in both halves show the generation of plasma at the contact points or the locations opposite to it at the planar counter electrode in both halves of the voltage cycle; which have been identified as surface microdischarges in previous work [9]. These microdischarges are generated at the positions, where laterally propagating SIWs meet and the local charge density is maximum.

The corresponding 2D images in Figure 3 show that a discharge is initiated uniformly at the apex of each dielectric structure at 42 and 93 µs. The emission is relatively hard to see in Figure 2 (b). It is caused by cathode directed positive streamers. The following image at 44 µs shows brighter emission caused by the FMD (ICCD peak counts increase ≈ 20 times compared to 42 µs). However, the filaments are brighter above the central pellets. The following images at 45, 46 and 47 µs show that the higher intensity spreads with time towards the edges of the array of dielectric hemispheres until all the filaments achieve a comparable intensity at 47 µs. This is followed by a decrease of the emission of the central filaments from 48 – 51 µs. At 49 µs the filaments located at the outermost pellets have a high intensity and at 54 µs only the outermost filaments are visible.

A similar behavior is observed during the phase of positive polarity of the driving voltage waveform. The initial breakdown happens almost uniformly at 93 µs, but it becomes much brighter in the central region at 95 µs. At 96 µs, the intensity is high above almost all the dielectric hemispheres, even close to the edge of the array. Due to the propagation of SIWs towards the contact points, the discharge at the contact points is also visible at 96 µs. This is different from the negative polarity, because the acceleration of electrons due to the polarization of the dielectric over the surface of the pellets in the positive half is much stronger than the acceleration of electrons over the flat dielectric. Again, the emission intensity of the microdischarges in the central region decreases from at 97 µs until it almost vanishes at 99 µs.

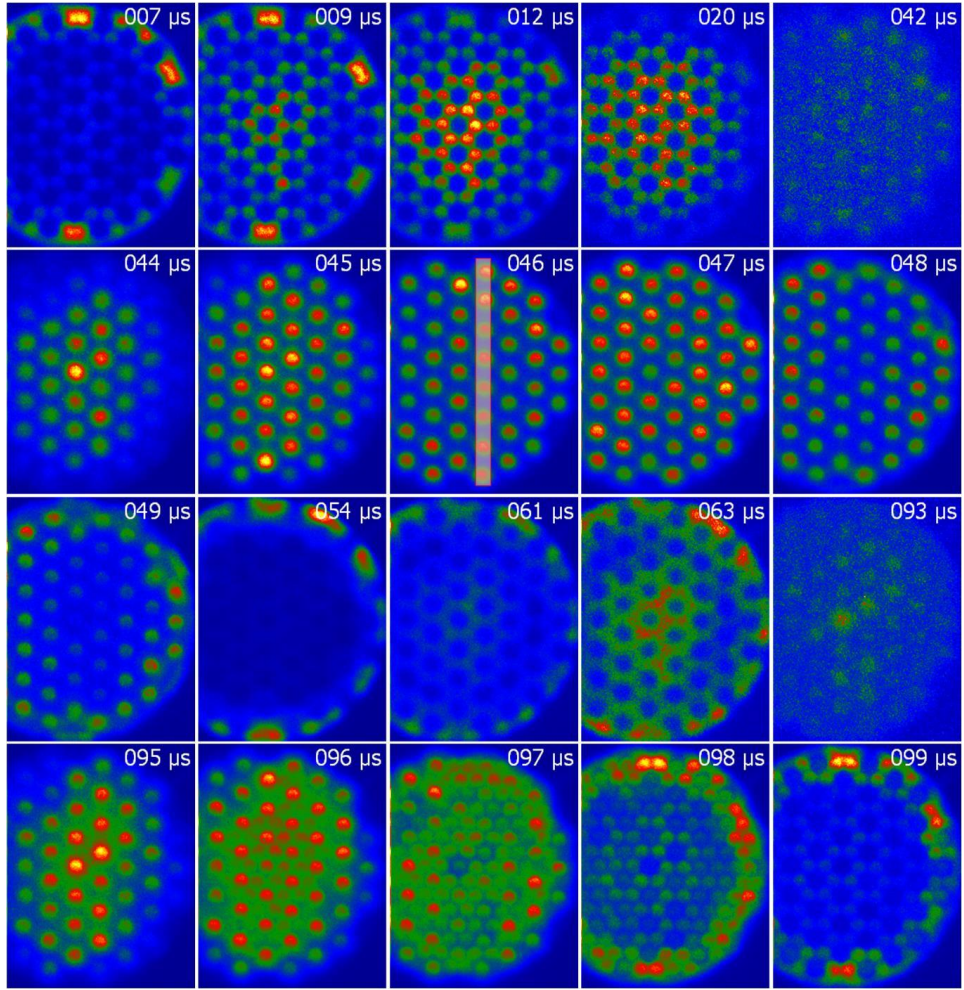

*Figure 3: Selected full view 2D images corresponding to an applied voltage amplitude of 1.97 kV. Each image corresponds to a specific time of acquisition according to Figure 2. The image corresponding to 46 µs also shows the position of the time resolved emission plotted in Figure 2.*

For both polarities, a second discharge pulse occurs in the form of a surface microdischarge at 63 and 9 µs, for the negative and positive polarity, respectively. The horizontal line in Figure 2 (b) also shows that the secondary discharge is generated between the apexes of two dielectric structures. The intensity of the discharge again increases in a wave like manner from the center towards the edges for both polarities e.g. during 9 – 20 µs for the positive polarity. The wave propagation speed is much slower for the second pulse compared to the first one as well as the intensity depicted from Figure 2 (a).

Geometrically this plasma source consists of an array of meso-micro scale cavities. Multiple segmented discharges are formed in each cavity. In this work, a significant number of pellets is used which allows to study the modulation of the plasma emission inside these cavities.

The results show that the discharge is initiated almost uniformly at all lateral locations. However, the plasma emission intensity increases in the center and propagates towards the edges in a wave like manner. Such wave like behavior has been observed before in microscopic array discharges [19] [20] and surface microdischarge arrays [18]. The non-uniform ignition was associated to the electric field distribution near the power connection. The wave like propagation was associated to the photoionization-based electron seeding from the brighter regions into the adjacent cavities. Based on the time resolution of 1 µs and the spatial resolution limited by the number of cavities, the wave propagation speed is estimated to be ≈ 10 and 12.5 km/s for the negative and positive polarities, respectively. The observed propagation speed of 10 km/s and 12.5 km/s for the two polarities is comparable to the estimated speed of 3 km/s [19], 5 km/s [20] and 30 km/s [18] in previous works.

In our case, the electric field distribution above the apexes of the dielectric structures seems to be uniform due to the uniform ignition of the initial discharges. The increase in the emission in the center could be explained as follows. The photo ionization is higher in the central cavities compared to the edges due to the higher number of seed photons from the adjacent cavities. Cavities located at the edge of the array have a fewer number of neighboring structures and, thus, the photo ionization seeding of electrons is less pronounced at the edge compared to the center. This results in a lower initial emission intensity at the edge of the array. Although cavities located further towards the center of the array, but still close to the edge, have the same number of neighboring cavities as those located in the center, some of these neighboring structures are characterized by a lower initial plasma density, because they are located at the edge and do not experience efficient photo ionization seeding themselves. Thus, the initial emission intensity is still weaker at cavities located close to the edge and there is a continuous increase towards the center of the array. This explains the higher emission intensity in the central cavities in the beginning.

The filamentary micro discharge (FMD) extinguishes once the dielectric at that location is charged. As the FMD intensity increases first in the center, the dielectric at these locations are also charged first and consequently, the discharge is extinguished first in the center. The photo ionization seeding leads to the wave like propagation of the emission intensity from the center to the edges. These observations can be identified due to the high number of pellets and the regular packing arrangement. This behavior has possible consequences for the length of the discharge as it results in a change in discharge ignition, dynamics and possibly efficiency based on the discharge dimensions. The extent of the interaction between multiple cavities would depend on the pellet size, gas and pellet dielectric constant.

The time resolved square of the measured current and the plasma emission both have maxima at ≈ 48 μs and ≈ 99 μs. There is a strong correlation with the emission structures shown on a logarithmic scale in the right column of Figure 2. Both full view (top) and zoomed (bottom) plots show that in both half cycles, the first emission structure is generated at the apex of each dielectric structure, i.e. at the location of minimum gap between the dielectrics.

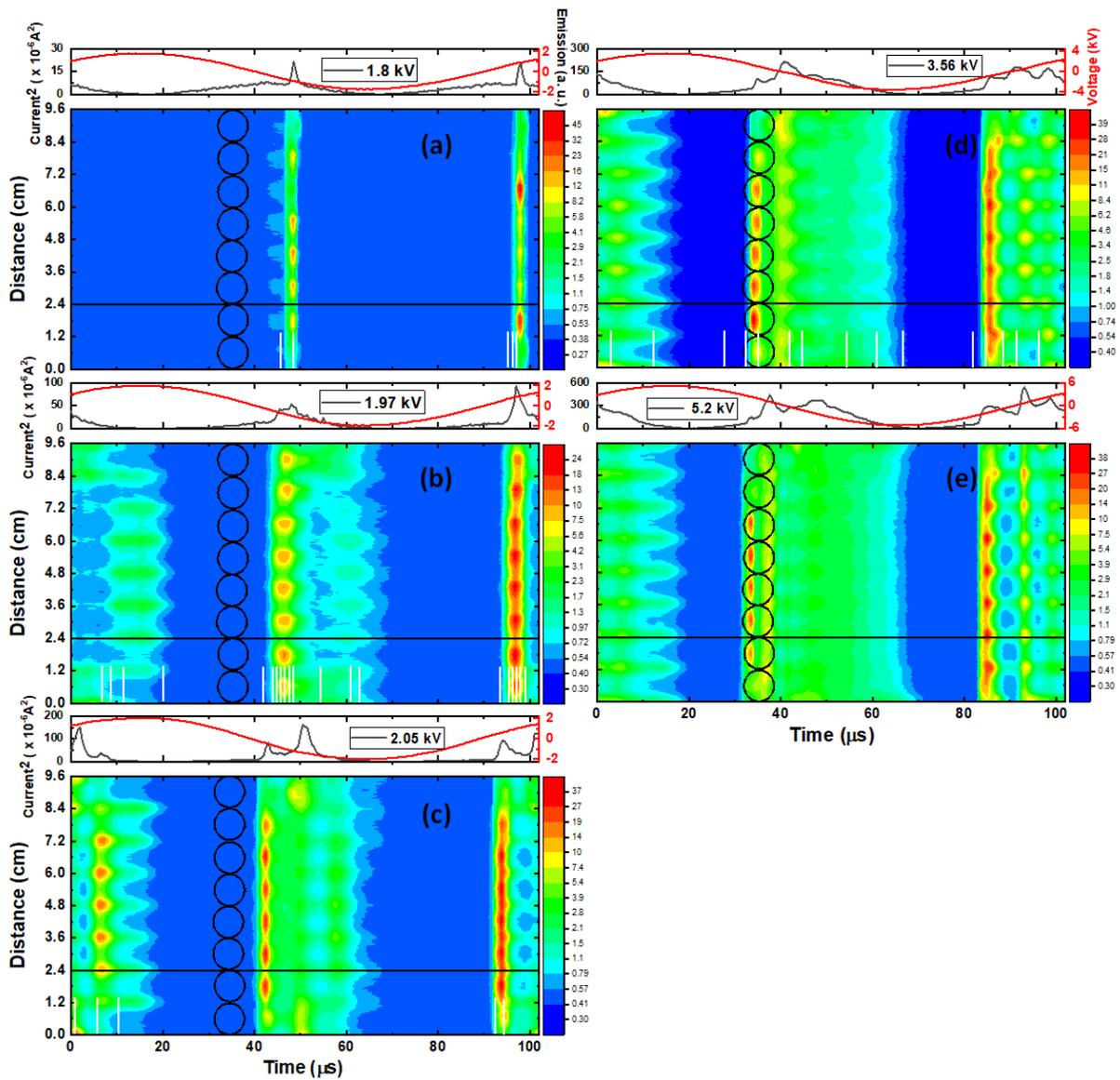

*Figure 4 Square of the current and driving voltage during the complete voltage cycle (top) and spatio-temporally resolved emission from the He ($3^3S_1$) state for different applied voltage*

*amplitudes of (a) 1.8 kV, (b) 1.97 kV, (c) 2.05 kV (d) 3.56 kV and (e) 5.2 kV. The acquisition time of the corresponding 2D emission images shown in Fig. 5, is indicated with white lines. The generator power was increased continuously from (a) to (e).*

Figure 4 shows the spatio-temporally resolved emission taken at the position A of the full view of the discharge (in Figure 1 (a)) at different applied voltage amplitudes ranging from 1.8 to 5.2 kV. The vertical axis shows the lateral spatial position [indicated in Figure 3], and the horizontal axis shows one complete cycle of approximately 102 µs. All data have been plotted on a logarithmic scale to clearly observe the multiple pulses of plasma emission simultaneously. Each space-time contour plot is correlated with the time resolved square of the current as well as the driving voltage waveform (top). The squared current is related to the power deposition and linked with the modulation in plasma emission (bottom). Figure 5 shows the corresponding full 2d spatially resolved plots at different times within one period of the driving voltage waveform at different voltage amplitudes with the exception of 1.97 kV, which was already shown in figure 3. One effect of increasing the applied power is an earlier initiation of the discharge in both halves. Therefore, the first emission maximum in the negative half appears at ≈ 48, 46, 42, 35, 33 µs for the applied voltage of 1.8, 1.97, 2.05, 3.56 and 5.2 kV, respectively.

The time-resolved emission from the discharge for an applied voltage amplitude of 1.8 kV is shown in Figure 4 (a) and the corresponding 2D images are shown in Figure 5 (a). This scenario corresponds to the lowest power at which a stable and uniform plasma could be generated. At the lowest power, only a single emission structure is generated in each half cycle corresponding to the FMD at the apex of each hemisphere. There are eight emission maxima corresponding to the apexes of eight hemispherical pellets at the positions of the minimum gap between the dielectrics. The intensity is not uniform due to the wave-like propagation resulting from the interaction between the adjacent microdischarges discussed above.

Figure 5 (a) shows that the discharge is generated unformly at the center of all pellets by positive streamers (PS) at ~ 42 and 96 µs corresponding to the negative and positive half cycles, respectively. At the lowest power, only a single emission structure is generated in each half cycle, due to a PS followed by a FMD at each dielectric structure and in each half cycle. Later the emission propagates as a wave from the center towards the edges indicated with images at ~ 49 and 98 µs. The discharge is still generated at the center of the dielectric structures, representative of the FMDs. Under these conditions, additional microdischarges are not formed as the voltage drop across the gap and the surface charging are not high enough to cause further ionization

Increasing the generator power leads to the results shown in figure 4 (b). The number of discharge structures increases to two, as discussed in detail in Figure 2, i.e. a SMD follows the FMD and the SIW. Further increasing the generator power results in a driving voltage amplitude of 2.05 kV and the number of emission pulses in each half cycle increases to three, i.e. a second SMD peak occurs after the first SMD peak. The emission structures are much sharper in space and time for the positive half compared to the negative half, because the positive streamers moves towards/along the structured dielectric surface at the grounded electrode. Due to the structured surface the different discharge phases are much more localized compared to the phase of negative polarity, where the streamer propagates towards/along the planar dielectric surface at the powered electrode and no apexes/contact points exist. Although the dielectric thickness is uniform at the powered electrode, the plasma emission shows stronger emission at positions opposite to the contact points below. This is because of the spatially non-uniform electric field generation due to the polarization of the dielectrics and its structured surface at the grounded electrode.

Figure 5 (b) shows the 2D images during the positive half of the voltage cycle corresponding to a voltage amplitude of 2.05 kV, where three emission structures are generated. It shows that during the first emission structure the discharge is initiated at the center of each dielectric structure at 92 µs, representative of a FMD. However, the emission is brightest for the central dielectric pellets. At 94 µs, the FMD during the same discharge structure transforms into a SMD as well as the discharge intensity propagates further radially. The 2D image at 1 µs shows that the second emission structure is generated as SMD, and the overall emission is brighter at the edges compared to the center. During the first discharge pulse, the wave did not propagate till the edge of the reactor and the propagation cycle was completed during the second excitation structure. For the third excitation structure, the plasma is generated as SMD and the wave propagation starts again from the center represented by the image at 6 µs. At 12 µs, the discharge wave is still propagating radially and simultaneously the intensity is reduced from the center. The behavior could be explained as follows. The duration of the discharge generation at a specific location depends on the dielectric charging which is independent of the wave propagation in the reactor, which depends on the interaction between the adjacent cavities.

A further increase of the generator power results in higher voltages ranging from 3.56 to 5.2 kV and in an increase of the number of emission pulses in both half cycles. The additional emission maxima correspond to additional SMDs. In the positive half, four pulses are clearly visible and the spatial emission structures are also clearly identified. In the negative half, at least three pulses are visible. The pulses and spatial structures are generally blurred during the negative half.

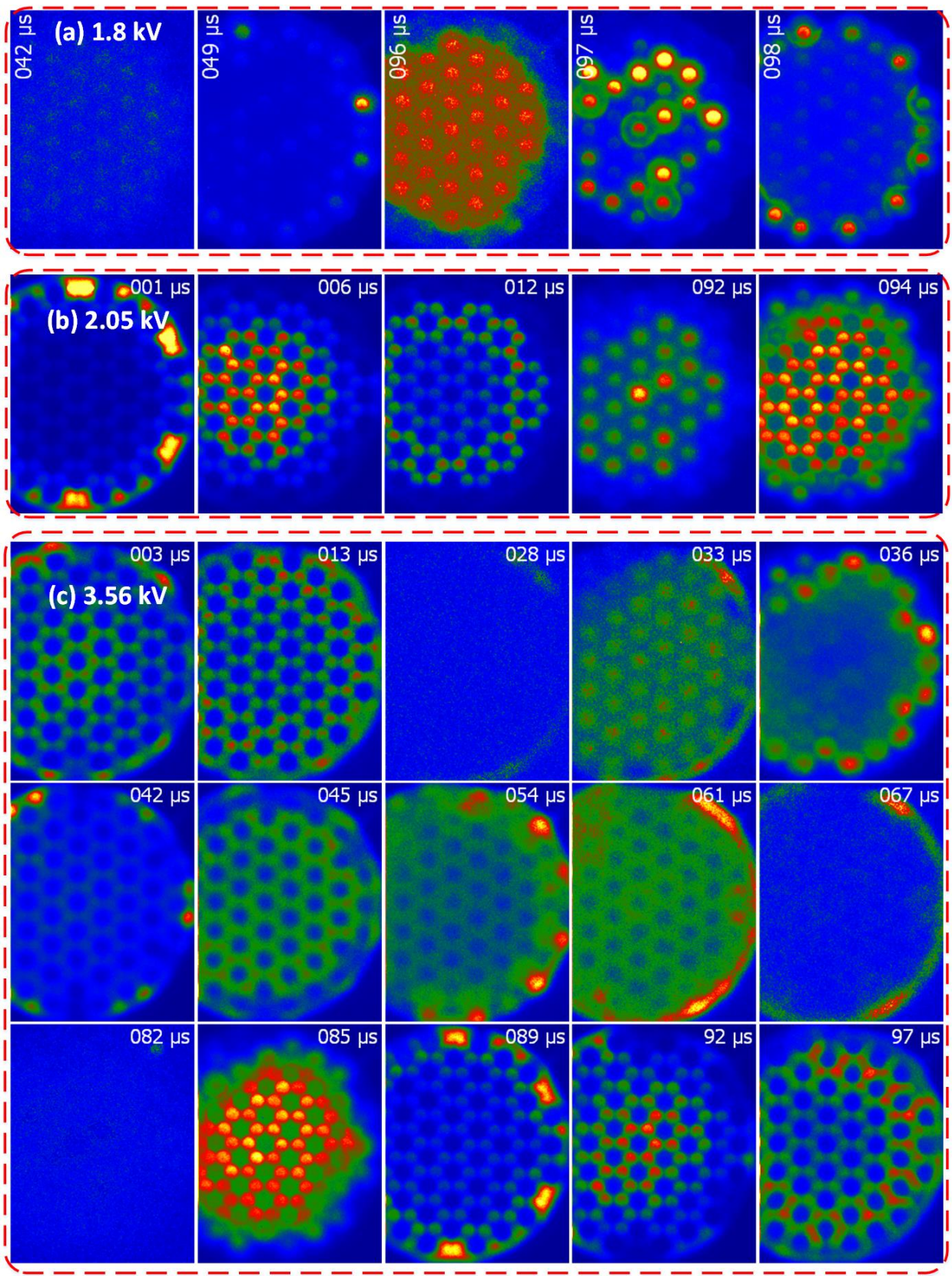

*Figure 5: Selected full view 2D images corresponding to an applied voltage amplitude of (a) 1.8 kV, (b) 2.05 kV and (c) 3.56 kV. Each image corresponds to a specific time of acquisition consistent with the corresponding space-time plots in Figure 4. The time of the acquisition is represented in Figure 4 as white lines.*

At high voltage amplitudes of 3.56 and 5.2 kV the plasma emission does not extinguish during the entire voltage cycle, as shown by the 2D images at 28, 67 and 82 µs. They indicate that after the discharge is extinguished at the center of the dielectric structures, it remains as a ring (at 28 and 67 µs) for both polarities or as a few dots (at 82 µs). The 2D image at 33 µs shows that the discharge is generated as positive streamer almost uniformly, in addition to the ring structure at the edge. Later, it propagates as a wave towards the edge represented by the images at 36 µs (during the negative half) and 85 and 89 µs (during the positive halves). The discharge generates several additional waves as shown during the positive half. The images at 92 and 3 µs represent the examples of the first phase of the wave propagation with higher brightness in the center. The images at 97 and 13 µs depict the second phase with reduced intensity in the center and higher intensity at the edges. At this time, the wave has propagated till the edge and the discharge has started extinguishing from the center.

At the lowest power, the duration of the filamentary microdischarges is ~ 3 µs. The duration of this emission structure initially increases as a function of power up to the point, where the driving voltage amplitude is 1.97 kV. With the further increase in applied voltage to 5.2 kV, the duration of the filamentary microdischarge structure decreases. This is because of the increase of the speed of the positive streamer with voltage. The filamentary microdischarge is able to charge the dielectric faster which results in faster extinguishing of the FMD. For both polarities, the electrons are almost simultaneously accelerated radially on the dielectric surface so that SIWs are formed. For the positive half, the apex of the pellets is charged and the the acceleration of the SIW towards the contact points is much stronger. Also, the emission during the acceleration towards the contact points increases such that for an applied voltage amplitude of 3.56 and 5.2 kV, the initial filamentary microdischarge cannot be noticed due to faster acceleration and lower relative emission intensity and it appears as the discharge is ignited at the contact points.

Due to the effect of interaction between the adjacent microdischarge and cavities, the wave like propagation of the emission from the center to the edges appears multiple times in each half cycle. The presence of this wave like propagation is indicated by brighter structures in the center or the edges and an arch like structure in terms of time. The wave-like motion speed increases as a function of the voltage amplitude from ≈ 12.5 to 25 km/s (for the positive half). Also, it is higher for the first emission structure and decreases for the successive emission structures at each power.

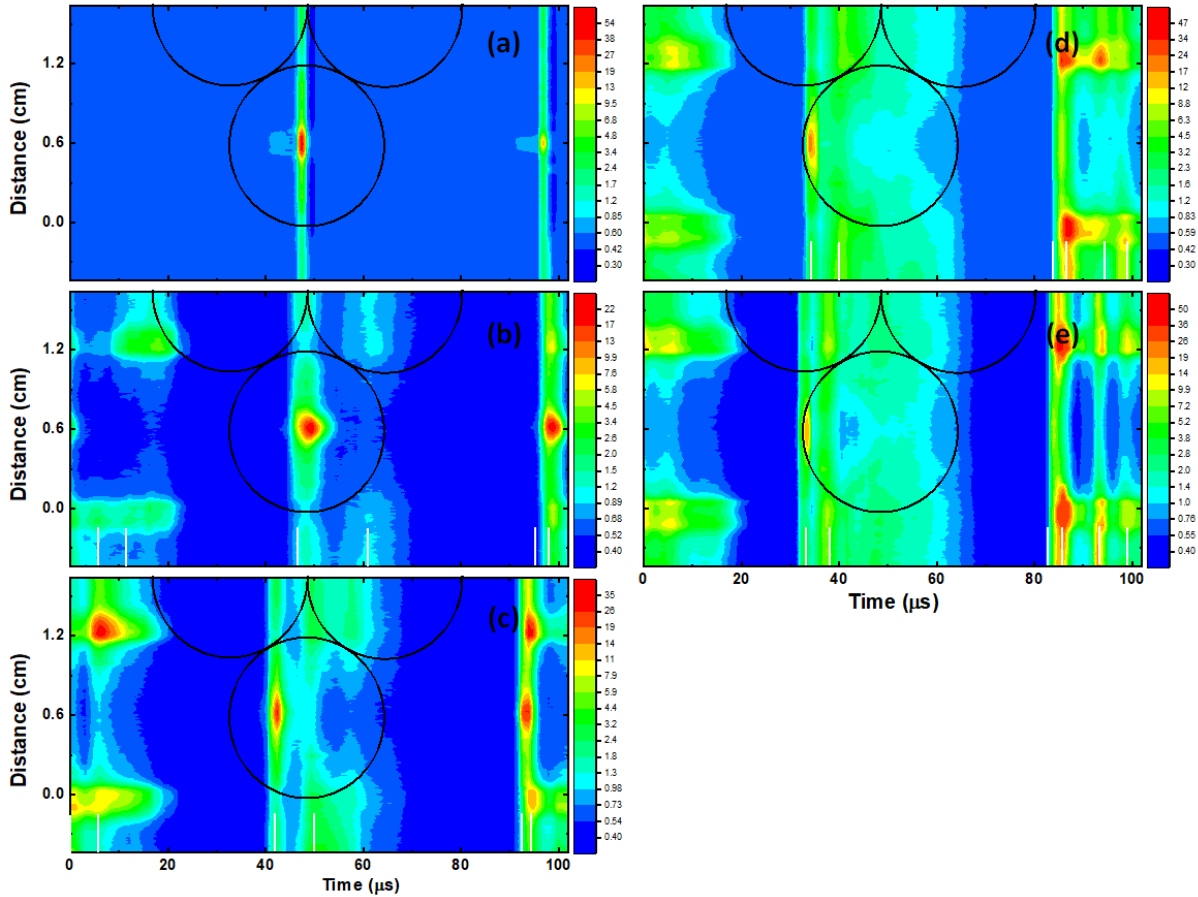

*Figure 6: Zoomed spatio-temporal plots of the emission from the He ($3^3S_1$) state for voltage amplitudes of (a) 1.8 kV, (b) 1.97 kV, (c) 2.05 kV (d) 3.56 kV and (e) 5.2 kV corresponding to position B in Figure 1 (d). The corresponding 2D emission images are shown in Fig. 7.*

Figure 6 shows the spatio-temporally resolved emission from the position B of the zoomed view of the discharge (in Figure 1 (a)) for multiple applied voltages ranging from 1.97 to 5.2 kV. The vertical axis shows the vertical distance B [indicated in Figure 1 (a, d)], and the horizontal axis indicates one complete cycle of approximately 102 µs. The space time contour plots in Figure 6 correlate with selected 2D time resolved images in Figure 7. Figure 6 shows higher spatial resolution compared to the Figure 4, to observe the plasma formation over the surface of a single pellet and the contact point. The number of emission structures in each half cycle is, therefore, similar in Figures 4 and 6. Figure 6 (a) shows the space time contour at the position B at the lowest power of 1.8 kV. It shows that during the initial PS phase faint emission is only observed in the center of the pellets whereas during the FMD phase the emission increases and expands till the edge of the pellet.

At 1.97 kV (Figure 6 (b)), at least one more excitation structure is generated in each half cycle. With the increase in applied voltage from 1.97 kV to 5.2 kV (Figure 6 (b) – (d)), the number of excitation structures increases further. The axial position shows that all the successive emission structures (after the first one) are generated at the contact points (SMD).

Previous results showed that the FMD is followed by an excitation structure due to radial acceleration of electrons identified as SIW. Here, for both polarities, the electrons are almost simultaneously accelerated radially due to a component of the electric field tangent to the surface. However, the radial acceleration of electrons generates lower emission compared to the FMD in the center of pellet. The whole duration of this radial acceleration is too short to be observed clearly with 1 µs resolution.

This phase is, therefore, investigated in Figure 7 with 0.1 µs time resolution, for the voltage amplitude of 3.56 kV. It shows that emission starts at 84.5 µs. At 84.8 µs, a ring like structure is formed near the apex representing the radial electron acceleration or SIW. Simultaneously, the whole dielectric structure surface is covered by weak emission. At 85 and 85.4 µs, emission is generated by both the SIW ring and at the contact points. During this phase, the diameter and emission intensity in the ring increases along with the intensity at the contact points. Simultaneously, the whole dielectric structure shows weaker emission. At 86 and 86.5 µs, the emission at the contact point increases while the emission of the SIW ring decreases before it reaches the contact point. The behavior could be explained as follows.

A ring like structure is formed near the apex representing the SIW. However, the whole dielectric structure is bright at 84.8 µs and SMDs are formed in the successive photos. This indicates the presence of energetic electrons spread out over the entire dielectric surface with different energies at different lateral positions. This is caused by the lateral propagation of the positive streamer head across the dielectric surface, which attracts electrons. At the contact points of adjacent dielectric structures, the local charge density is expected to be maximum due to the merge of several SIWs propagating across the adjacent dielectric hemispheres. Thus, strong electric fields are generated at the contact points and significant emission is present. The charge accumulation at the contact points is expected to decelerate the approaching SIWs.

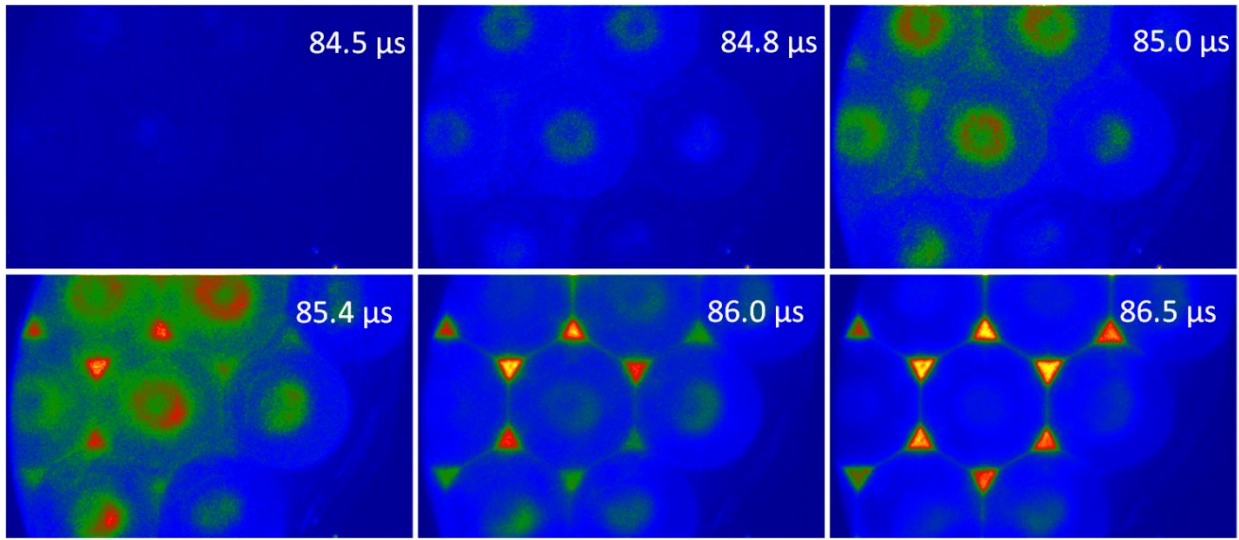

*Figure 7: Selected 2D images of the zoomed region corresponding to the black region of interest indicated in Figure 1 (a) for applied voltage amplitudes of 3.56 kV. The images correspond to the space time contours shown in Figure 6.*

The emission due to the FMDs decreases as a function of the driving voltage amplitude relative to the SMD. This change of the discharge structure as a function of voltage at different phases is illustrated by Figure 8, which shows zoomed 2D images taken at different times within one period of the driving voltage waveform for applied voltage amplitudes of 1.97, 2.05, 3.56 and 5.2 kV.

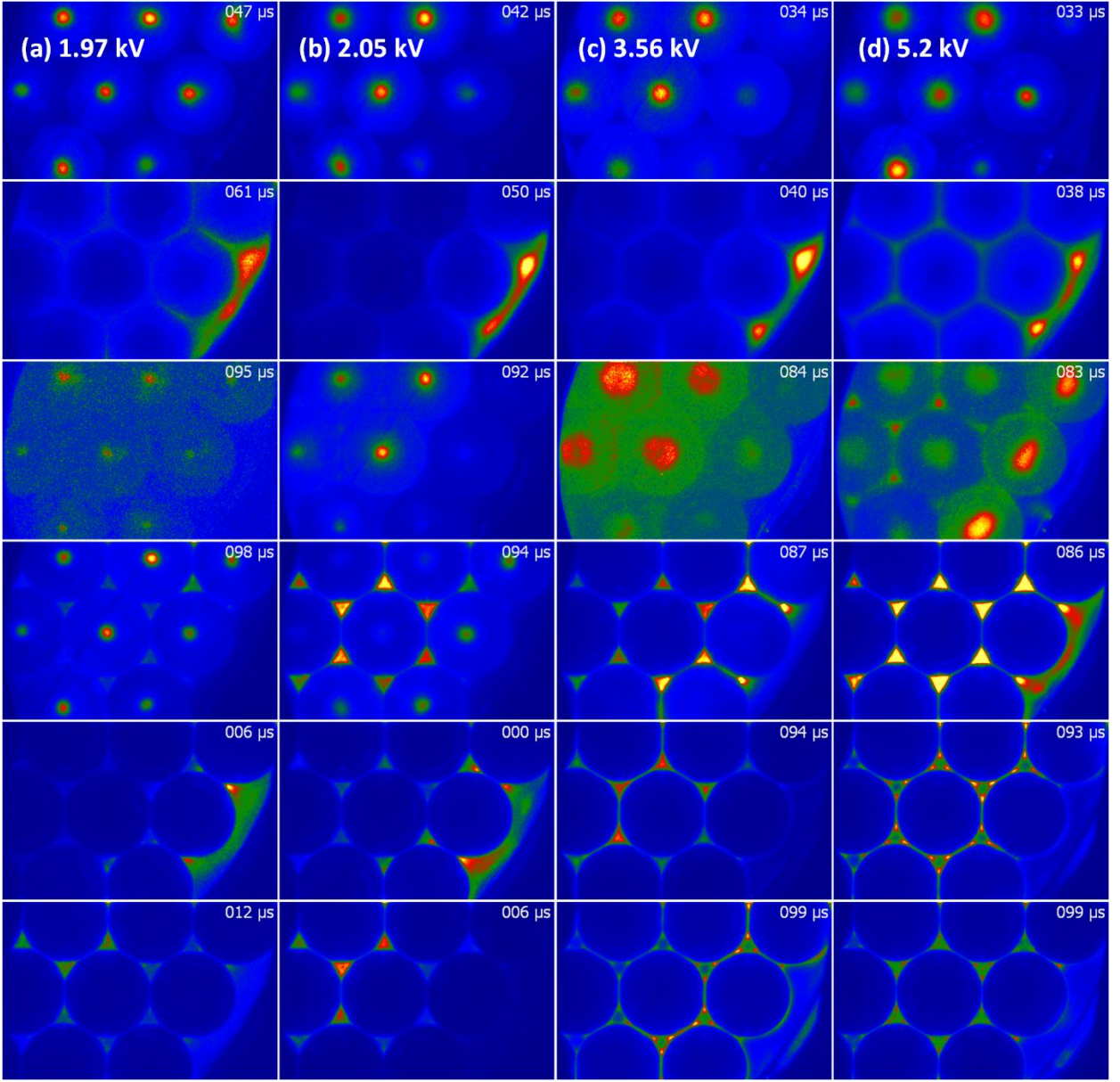

*Figure 8: Selected 2D images of the zoomed region corresponding to the black region of interest indicated in Figure 1 (a) for different applied voltage amplitudes of 1.97, 2.05, 3.56 and 5.2 kV at different times within one period of the driving voltage waveform. The images correspond to the space time contours shown in Figure 6.*

Each column corresponds to a specific voltage amplitude. The time of each image is indicated by a line in the corresponding space-time plots shown in Figure 6. The first two rows correspond to the negative half cycle when the discharge is predominantly generated over the flat dielectric. The first row show images taken 3 µs after the plasma ignition during the negative half cycle, which shows that the plasma is ignited as a FMD at each power.

The second row represents the time of maximum emission of the second emission structure. It shows that a discharge is generated over the flat dielectric at locations opposite to the contact points due to the electric field distribution. The discharge is, however, strongest at the edges due to the wave like propagation discussed above.

The third row shows images taken 3 µs after the plasma ignition during the positive half cycle. It indicates that plasma is generated by a PS followed by the FMD. Electrons are accelerated along the dielectric surface due to stronger electric fields and due to the curvature of the dielectric. At the lowest voltage, only the emission at the center of the dielectric structure is visible indicating a PS or FMD. At 2.05 kV, emission is also generated over the surface of the dielectric. At 3.56 kV, the discharge has a ring like shape which is indicative of the SIW. Although, the ring is located relatively close to the apex of the dielectric structure, there is still simultaneous emission at the contact points. At 5.2 kV, the emission at the contact points is visible in addition to emission on the surface and at the apex. This shows that with the increase in voltage amplitude, the discharge attains the PS to FMD to SIW and SMD in a shorter time.

The fourth row represents the time of maximum emission for the first emission structure in the positive half cycle, during which the discharge transforms from a PS to a FMD to a SIW and finally to a SMD. At the lowest voltage of 1.97 kV, the emission is maximum at the apex of the dielectric structure with weaker emission at the contact point. At 2.05 kV, the emission is maximum at the contact point and weaker emission is observed at the apex. At 3.56 and 5.2 kV, the emission is only generated at the contact point. This shows that with the increase in voltage amplitude, the discharge attains the FMD to SMD transition during a shorter time.

The fifth and sixth row represents the time of maximum emission for the second and third emission structure in the positive half cycle, respectively. The second and third emission structures represent the generation of SMDs at the contact points and the surface microdischarges are visible at the contact point for all voltages. The sharpness and brightness of the SMD emission increases with the voltage amplitude. For 3.56 and 5.2 kV, the SMD is segmented into three sharp microscopic structures at the three edges. This could be explained by the stronger polarization of the dielectrics at the contact points at higher applied voltages and by the enhanced local plasma density, which allows the plasma to penetrate deeper into narrow structures at high voltage due to the decreased sheath width. This is consistent with the previous simulation study of discharges inside dielectric pores, where discharge formation inside the pore becomes pronounced at higher voltage amplitudes [12].

This is the first observation and confirmation of the discharge formation in the void between the dielectrics. As the discharge is predominantly generated in the void between the dielectrics as a surface discharge at these higher voltages and the effect of the microdischarge formation seems to play a dominant role under these conditions, the understanding of the surface discharge segmentation is important to model these discharge conditions correctly.

There are six contact points around each dielectric structure. Selected images in Figure 3 (e.g. at 10 and 20 µs) show that one microdischarge of similar intensity is produced at each contact point around a dielectric simultaneously. This shows the potential of the regularly packed dielectrics for producing an atmospheric pressure plasma over a large area. These results also show that the SIW phase is merged with the first current pulse in the presence of the relatively small dielectric structures used here instead of the presence of separate pulses such as observed for larger sized dielectric structures. These dynamics can be controlled by the size and/or the curvature of the dielectric structures. The observed results confirm previous predictions of models experimentally. These models predicted that the size of the dielectric pellets is an important parameter that controls the plasma dynamics as the basis for process optimization [26]. The observed wave-like behavior of the plasma generation affects the discharge duration and dimensions. This is expected to play an important role for the efficiency of different applications such as gas cleaning. The interaction between multiple cavities could be a crucial factor for the performance of commercial devices to achieve higher throughputs, e.g. higher gas processing flowrates. The interaction between multiple cavities is expected to depend on the pellet size, shape, gas and pellet dielectric constant. The effects of these parameters on the plasma generation still remain to be studied. Moreover, the consequences of the interaction between the cavities on different plasma processing application should be investigated in the future. The cavities between the dielectric pellets are very important as the plasma-catalyst interaction happens at these locations. Also, the behavior is representative of the discharge formation in surface pores which have been modelled, but have not been investigated experimentally. It is observed that at low voltages the surface discharge in the cavity has the shape of the cavity. However, at high voltages multiple microscopic structures are formed within the cavities. Due to the intense plasma generation on these microscopic scales the plasma chemistry could be very different from that within filamentary discharges in conventional DBDs.

## 4. Conclusions

Plasma dynamics in a patterned dielectric barrier discharge with regularly arranged pellets embedded into the surface of the powered electrode is investigated in helium gas and in the presence of a sinusoidal driving voltage waveform with different amplitudes of 1.8 – 5.2 kV. The results show that plasma is initiated as cathode directed streamers followed by filamentary microdischarges at the positions of the minimum electrode gap at the apex of the embedded dielectric structures similar to conventional DBDs. This is followed by the instant radial acceleration of electrons along the dielectric surface. Multiple discharge pulses are generated due to surface micro discharge formation at the contact points, whose number depends on the driving voltage amplitude. Zooming into the cavities at the contact points, a given surface microdischarge is observed to be segmented into smaller microscopic structures whose emission intensities increase with the applied voltage. The interaction between adjacent filamentary and surface microdischarges located at different dielectric structures generates a wave-like emission intensity propagation from the center of the dielectric array to the edges caused by electron seeding due to photoionization caused by plasma generation at adjacent dielectric structures.

**Acknowledgement**

The authors would like to acknowledge the support of DFG project number 432514770 and from the SFB 1316 (project A5).